\documentclass[final,1p,times,twocolumn,sort&compress,fleqn]{elsarticle}

\usepackage{graphicx}

\usepackage{amssymb}
\usepackage[small,bf]{caption2}
\usepackage[bookmarksopen,bookmarksnumbered,colorlinks,linkcolor=blue,anchorcolor=blue,citecolor=green]{hyperref}
\usepackage[all]{hypcap}

\journal{Physics A}

\begin{document}

\begin{frontmatter}


\title{Characteristics of Real Futures Trading Networks}
\author[scsfdu,shfe]{Junjie Wang}
\ead{wangjunjie@fudan.edu.cn}
\author[scsfdu,skliipfdu]{Shuigeng Zhou\corref{col1}}
\ead{sgzhou@fudan.edu.cn}
\author[dcsttju]{Jihong Guan\corref{col1}}
\ead{jhguan@tongji.edu.cn}

\cortext[col1]{Corresponding author}
\address[scsfdu]{School of Computer Science, Fudan University, Shanghai 200433, China}
\address[shfe]{Shanghai Futures Exchange, Shanghai 200122, China}
\address[skliipfdu]{Shanghai Key Lab of Intelligent Information Processing, Fudan University, Shanghai 200433, China}
\address[dcsttju]{Department of Computer Science and Technology, Tongji University, Shanghai 201804, China}

\begin{abstract}
Futures trading is the core of futures business, and it is
considered as one of the typical complex systems. To investigate the
complexity of futures trading, we employ the analytical method of
complex networks. First, we use real trading records from the Shanghai
Futures Exchange to construct futures trading networks, in which
nodes are trading participants, and two nodes have a common edge if
the two corresponding investors appear simultaneously in at least
one trading record as a purchaser and a seller respectively. Then,
we conduct a comprehensive statistical analysis on the constructed
futures trading networks. Empirical results show that the futures
trading networks exhibit features such as scale-free behavior with
interesting odd-even-degree divergence in low-degree regions,
small-world effect, hierarchical organization, power-law
betweenness distribution, disassortative mixing, and shrinkage of both
the average path length and the diameter as network size increases. To the
best of our knowledge, this is the first work that uses real data to
study futures trading networks, and we argue that the research
results can shed light on the nature of real futures business.
\end{abstract}


\end{frontmatter}

\section{Introduction}
Since the works of Watts \& Strogatz~\cite{Watts1998} and Barab\'
asi \& Alberta~\cite{Barab1999} were published, complex networks, as
a new scientific area, have attracted a tremendous amount of research
interest~\cite{Newman2003,Albert2002,Boccaletti2006,Costa2007,Song2006}.
Complex networks can describe a wide range of real-life systems in
nature and society, and show various nontrivial topological
characteristics not occurring in simple networks such as regular
lattices and random networks. There are a number of frequently cited
examples that have been studied from the perspective of complex
networks, including the World Wide
Web~\cite{Huberman1999,Albert1999,Broder2000,Baraba2000}, the
Internet~\cite{Faloutsos1999}, metabolic networks~\cite{Jeong2000},
scientific collaboration networks~\cite{Newman2001,Barrat2004a},
online social networks~\cite{Subrah2008,Fu2008}, public transport
networks~\cite{Kurant2006,Su2006}, airline flight
networks~\cite{Guimera2004,Barrat2004a,LiW2004,LiuHK2007,Han2009}
and human language networks~\cite{Motter2002,Calderia2006,Zhou2008}.
Empirical studies on these networks mentioned above
have largely motivated the recent curiosity and concern about this
new focus of research so that a number of techniques and models have
been explored to improve people's perception of topology and
evolution of real complex
systems~\cite{Albert2000,Bianconi2001,Zhu2004,Gallos2005,Newman2005,Tummi2005,LiuJG2006,LiuJG2007}.
With growing of importance and popularity, complex network
theory has become a powerful tool with intuitive and effective
representations to analyze complex systems in a variety of fields,
including financial markets~\cite{Costa2008}.

In the literature, a number of papers have been dedicated to
studying financial markets from the perspective of complex networks.
The major difference among these works lies in the types of networks
to be constructed from financial data for characterizing the
organization and structure of financial markets. Some existing works
constructed stock networks whose connectivity is defined by the
correlation between any two time series of stock
prices~\cite{Mantegna1999,Bonanno2003,Onnela2003,Onnela2004,Pan2007}.
Some others established directed networks of stock ownership
describing the relationship between stockholders and
companies~\cite{Garlaschelli2005,Battiston2007}. Networks of market
investors based on transaction interactions between the investors
were also investigated. For example, Franke et al.~\cite{Franke2007} analyzed irregular
trading behaviors of users in an experimental stock
market, while Wang et al.~\cite{Wang2008} studied the evolving
topology of such a network in an experimental futures
exchange. Recently, Jiang et al.~\cite{Jiang2010} has investigated stock
weighted directed trading networks based on daily transaction
records through reconstructing the limit order book with real order
series from the Shenzhen Stock Exchange.

The study on financial investor networks can provide clues for revealing
the true complexity in financial markets, especially futures markets.
In a real futures market, the futures trading model serves as a
matching engine for executing all eligible orders from various market
participants, and the interactions among the participants form a
complex exchange network, which is termed as the \emph{futures trading
network} (FTN in short) in this paper. Simply put, a FTN consists of a
set of trading participants, each of which has at least one
connection of direct exchange action to another by futures
contract(s).

In this paper, we try to provide a comprehensive study on the
characteristics of futures trading networks established with genuine
trading data from the Shanghai Futures Exchange. To the best of our
knowledge, this is the first work that uses real futures trading
data for constructing networks. So we assume that the empirical results may
unveil to some degree futures trading behaviors and shed light on
the nature of futures markets.

The rest of this paper is organized as follows. The FTN construction
method is introduced in Section~\ref{sec:ftn-construction},
including trading dataset and the details of network construction.
Section~\ref{sec:ftn-properties} presents the empirical results of
FTNs. Section~\ref{sec:conclusion} concludes the paper and
highlights some future work directions.

\section{Construction of Futures Trading Networks}
\label{sec:ftn-construction} In this section, we first introduce the
real trading data from the Shanghai Futures Exchange, which will be used
for constructing futures trading networks, and then present the details
of the construction of futures trading networks.

\subsection{Dataset}
For constructing the futures trading networks, we use real trading records
from the Shanghai Futures Exchange, which is the largest one in China's
domestic futures market and has considerable impact on the global
derivative market. Trade records are generated by matching orders or
quotes from buyers and sellers according to a certain rule of
price/time priority (first price, then time) via the electronic
trading platform of the futures exchange. There are hundreds of
thousands of matching results reported from the exchange in one
typical trading day. We use a dataset involving all futures
commodities in the derivative market from July to September, 2008.
For the trading records, we use virtual and unique IDs for representing
the trading participants, and all other sensitive information is
filtered for privacy preservation reason.

\subsection{Network Construction}

\begin{figure}[!htp]
    \centering
    \includegraphics[width=\textwidth]{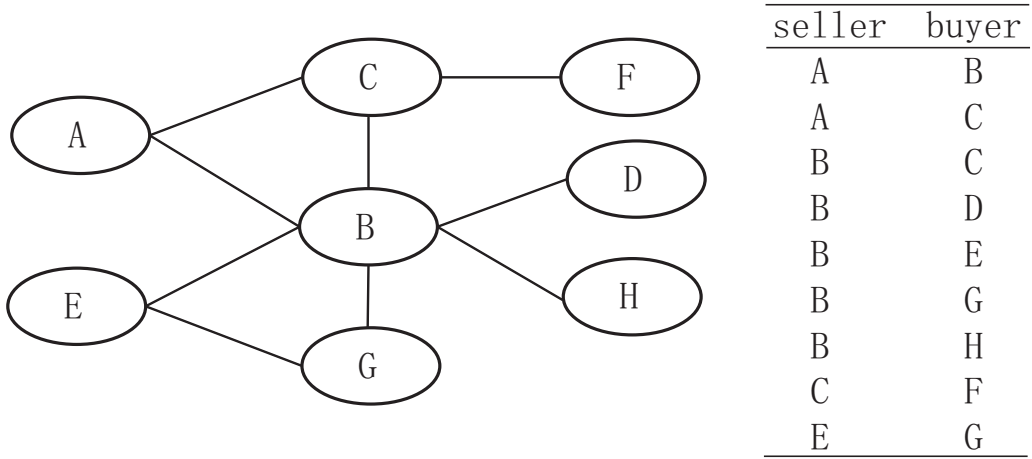}
    \caption{An example of the futures trading network (left) constructed from a dataset with nine trading records (right).}
    \label{fig:example-ftn}
\end{figure}

Since a trading record contains the IDs of both the buyer and the
seller, we are able to establish a futures trading network. The
construction process is as follows. A node represents a participant
in a trading record, and an edge, meaning a trade relation, is
established between two participants if their IDs appear
simultaneously at least once in one trading record. An example of
FTN that comprises nine records is shown
in~\autoref{fig:example-ftn}. Here, A-H are IDs of trading
participants. Each row in the right table of
~\autoref{fig:example-ftn} represents a trading record where the
seller and buyer columns are the IDs of seller and buyer.

Following the above process, we first create the largest network
with the whole dataset containing three months' trading records;
this network is denoted as FTN-all. Then, we construct three
sub-networks based on three subsets of the whole dataset above
according to the futures commodity classification of the Shanghai Futures
Exchange. These three sub-networks are termed as FTN-met, FTN-rub
and FTN-oil, which involve only futures commodity metal (concretely
including copper, aluminum, zinc and gold), natural rubber and fuel
oil, respectively. Thus, we have totally four FTNs, their statistic
information is given in~\autoref{tab:ftns}.

\begin{table}[!hbp]
    \captionstyle{indent}
    \caption{The statistic information of the four FTNs. FTN-all is based on the whole dataset
containing three months' trading records from July to September,
2008, FTN-met, FTN-rub and FTN-oil are three sub-networks of
FTN-all that correspond to three subsets that involve only trading
records of the futures commodity metal, natural rubber and fuel oil,
respectively.}
    \label{tab:ftns}
    \centering
    \begin{tabular*}{\textwidth}{@{\extracolsep{\fill}} cccc }
        \hline
        Network & Futures commodity & Number of nodes & Number of edges \\
        \hline
        \itshape FTN-all  & all commodities & 100994 & 8068676 \\
        \itshape FTN-met  & metal & 75262 & 3226119 \\
        \itshape FTN-rub  & natural rubber & 55828 & 3364557 \\
        \itshape FTN-oil  & fuel oil & 52208 & 1657268 \\
        \hline
    \end{tabular*}
\end{table}

\section{Properties of Futures Trading Networks}
\label{sec:ftn-properties} In what follows, we study the
characteristics of FTNs constructed above, our focus is on
topological features and dynamical properties. Extensive empirical
results are presented.

\subsection{Scale-free Behavior}
\label{subsec:scale-free}

\begin{figure}[h]
    \centering
    \includegraphics[width=\textwidth]{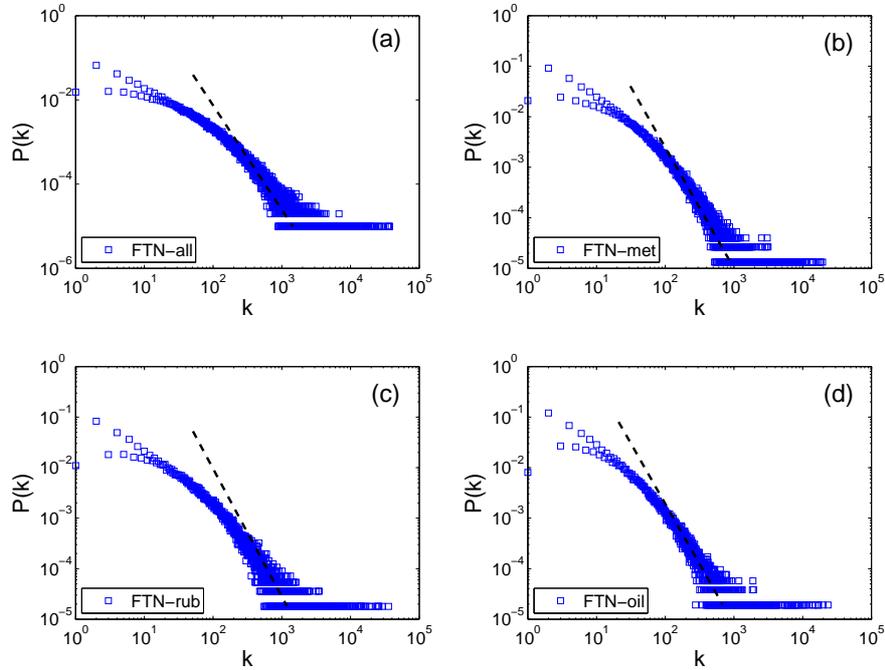}
    \caption{The degree distributions $P(k)$ of the four futures trading networks
    FTN-all, FTN-met, FTN-rub and FTN-oil. All the four fitting lines of the subplots
    have nearly identical slopes, and the scaling parameters $\lambda$ are 2.48, 2.39,
    2.53 and 2.39, respectively.}
    \label{fig:degree-distribution}
\end{figure}

The degree distribution {\it P(k)} is one of the most important
statistical characteristics of a network, and also one of the
simplest properties that can be measured directly. It is defined as
the probability that a random node in a network has exactly \emph{k}
edges. In many real complex networks, {\it P(k)} decays with $k$ in
a power law, following
\begin{equation} \label{e:DDF}
    P(k)\sim k^ {-\lambda}.
\end{equation}
A network that owns such a property is called a scale-free
network~\cite{Barab1999}. We plot the degree distributions $P(k)$ of
FTN-all, FTN-met, FTN-rub and FTN-oil in
\autoref{fig:degree-distribution}(a)-(d). Our
observation of the subplots implies that all the four FTNs are
scale-free networks. However, reaching a conclusion that all FTNs
are scale-free requires a deeper analysis by statistic techniques.

\begin{figure}[h]
    \centering
    \includegraphics[width=\textwidth]{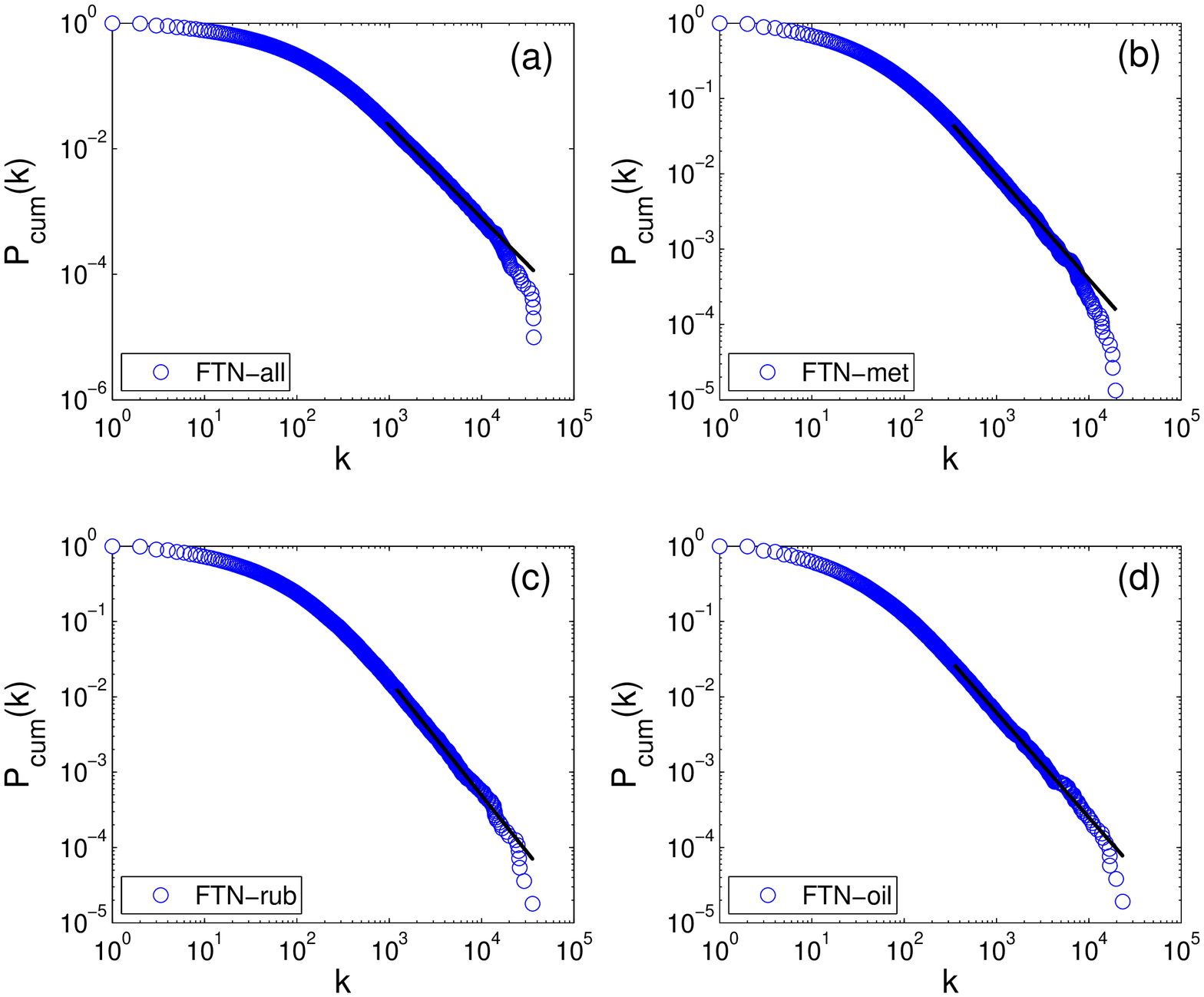}
    \caption{The cumulative degree distributions $P_{cum}(k)$ and their maximum likelihood power-law fits for FTN-all, FTN-met, FTN-rub and FTN-oil.
    The slopes of all fitting lines of the subplots are approximately similar, with values of 1.48, 1.39, 1.53 and 1.39, respectively.}
    \label{fig:cumulative-degree-distribution}
\end{figure}

To detect the existence of power-law property in FTNs,
we consider the cumulative degree distribution function defined as
$P_{cum}(k)=Pr(k' \geq k)$ because the power-law cumulative degree
distribution of a network means that its degree distribution obeys a
power law~\cite{Newman2005b}. According to Ref.~\cite{Clauset2009},
we calculate the complementary cumulative degree distributions of
FTNs for estimating the scaling parameter $\lambda$ and the lower bound
$k_{min}$ of power-law behavior based on the methods of maximum
likelihood estimators and the quantification of the distance between
two probability distributions. Furthermore, we conduct a
goodness-of-fit test by computing the distance between the empirical
distribution of FTN and the power-law model with the
Kolmogorov-Smirnov~(KS) statistic, and generating a {\em p}-value
that quantifies the plausibility of a power-law degree distribution
hypothesis for each FTN. The cumulative degree distributions
$P_{cum}(k)$ for all the four networks are presented in
\autoref{fig:cumulative-degree-distribution} with the slope
($\lambda$-1). \autoref{tab:dd-fit-test} gives the estimations of
$\lambda$, $k_{min}$, and the {\em p}-value of the goodness-of-fit
test for each power-law fit of FTN.

\begin{table}[!hbp]
    \captionstyle{indent}
    \caption{The estimations of the lower bound $k_{min}$ and the scaling parameter $\lambda$,
    and the goodness-of-fit tests of power-law behavior for the four futures trading networks.
    The {\em p}-values indicate that the degree distribution of each network is consistent with a power-law hypothesis.}
    \label{tab:dd-fit-test}
    \centering
    \begin{tabular*}{\textwidth}{@{\extracolsep{\fill}}  c *{3}{c}}
        \hline
        Network &\itshape $k_{min}$ &\itshape $\lambda$ &\itshape p-value \\
        \hline
        \itshape FTN-all & 926 & 2.48 & 0.225 \\
        \itshape FTN-met & 339 & 2.39 & 0.106 \\
        \itshape FTN-rub &1196 & 2.53 & 0.591 \\
        \itshape FTN-oil & 351 & 2.39 & 0.674 \\
        \hline
    \end{tabular*}
\end{table}

As listed in \autoref{tab:dd-fit-test}, the scaling parameters
$\lambda$ of the four FTNs have nearly similar values of 2.48, 2.39,
2.53 and 2.39 for FTN-all, FTN-met, FTN-run and FTN-oil,
respectively. Their {\em p}-values are 0.225, 0.106, 0.591 and
0.674, respectively, which provide sufficient evidence indicating
that the power law is a plausible hypothesis for the degree distribution
of each FTN. These results imply that highly connected nodes have
larger possibilities of happening and dominating the connectivity of
FTNs. Actually, these nodes correspond to a few active speculators
who issue numerous orders to the exchange, and consequently have
more opportunities to make deals with the others. Their trading
behaviors also account for the fact that FTNs dynamically expand in
accordance with the rule of preferential attachment by continuously
adding new nodes during the lifetime of a network.

In~\autoref{fig:degree-distribution}, we can also see the fact that
in low-degree regions, the probability of even degree (say $k$=2$n$)
is larger than that of odd degree ($k$=2$n$+1), which forms a
divergence. As the degree increases, the two branches gradually
reach to a convergence and the difference is averaged out. The
branch of even degree conforms to power law more than that of the
odd one. One reasonable explanation on this observation
is that most participants of low degree, entering the futures market
as speculators, cautiously make a few transactions; but worrying
about risk or loss, they do not hold the positions long, that is,
they open futures positions(buying/selling) and close them soon
(selling/buying). Certainly, there are some newcomers who are market
investors and prefer holding positions for long-term profits. To a
certain extent, due to the randomicity of matching counterpart during
trade execution, the two matches (open and close) for a speculator in
a complete open and close are very possibly involved in different
counterparts; thus two degrees (edges) will be added to the
corresponding node. This observation indicates that there are
few hedgers with long-term positions in real trading, which is
consistent with the fact of low delivery ratio in the futures
market.

\begin{figure}[!htp]
    \centering
    \includegraphics[width=\textwidth]{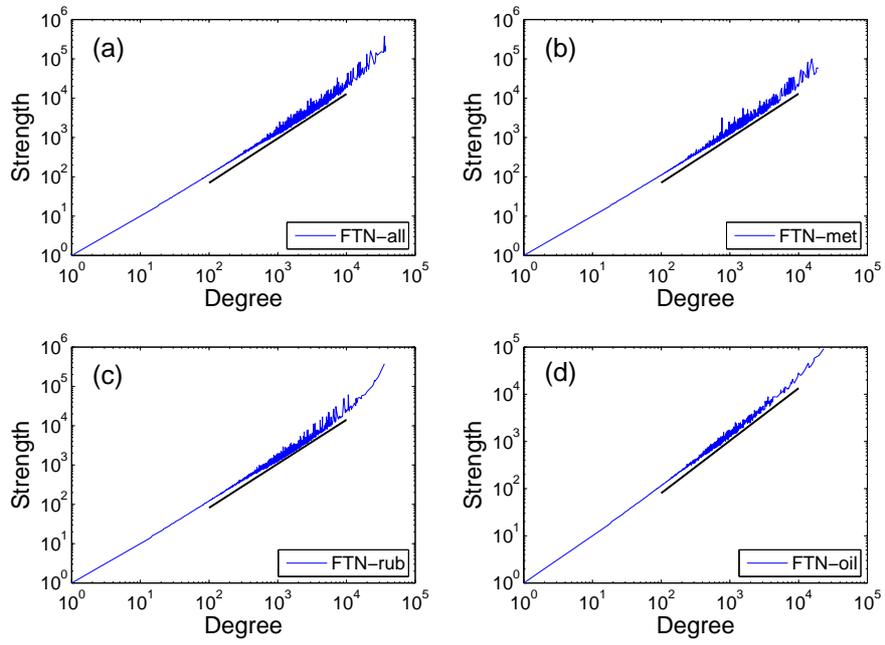}
    \caption{Node strength {\it S} versus degree {\it k} for FTN-all, FTN-met, FTN-rub and FTN-oil. Here, the slopes are nearly similar to 1.12$\pm$0.01.}
    \label{fig:strength-function}
\end{figure}

Now we consider weighted futures trading networks in
which our research focus is the relationship between strength and degree
of nodes. The strength-degree relation in air transportation
networks~\cite{Guimera2004,Barrat2004a,LiW2004,LiuHK2007,Han2009}
represents a power law, and various
models~\cite{Barrat2004b,Goh2005,WangWX2005,OuQ2007,ZhangZZ2007}
were proposed to explain the origin of the power-law correlation.
In this paper, the node strength is defined as follows. Suppose that
the weight of an edge between two nodes is the number of times they
appear in the same trade record, the strength of a node is defined
as the total weight of all edges connecting it.
\autoref{fig:strength-function} shows the relationship between the
average strength {\it S} of the nodes with degree {\it k} and {\it
k} in the four FTNs, which indicates that a nontrivial power-law
scaling $S\sim k^ \beta$ ($\beta$=1.12 for all the four networks)
exists, and demonstrates the fact that active market participants
get more active. Such a scale-free behavior being correlated to the
occurrence frequency of a trading participant provides another
justification for the power-law degree distribution of the networks.

\subsection{Small-world Effect}
For complex networks, the average path length and the clustering coefficient
are two important measures of small-world effect.

In a network, the average path length (say $L$) is defined as the
number of edges in the shortest path between any two nodes,
averaged over all pairs of nodes. It plays an important role
in transportation and communication of a network. We can obtain
$L=(2/n(n-1)) \sum _{i\geq j} d_{ij}$. The longest shortest path
among all pairs of nodes is called the diameter of the network.

The clustering coefficient $C_i$ of a node {\it i} is defined as the
ratio of the total number $e_i$ of edges that actually exist between
all its $k_i$ immediate (nearest) neighbors over the number $k_i
(k_i -1)/2$ of all possible edges between them, that is, $C_i=2e_i /
k_i(k_i-1)$. The clustering coefficient {\it C} of the whole network
is the average of $C_i$ over all nodes, i.e., $C=(1/n) \sum C_i$.
The clustering coefficient measures the probability that two
neighbors of a node are connected and reveals the local cliquishness
of a typical neighborhood within a network.

In recent empirical studies, many real systems show the small-world
effect by two crucial factors: the average path length and the diameter is
relatively small despite often the large network size, which grows
nearly logarithmically with the number of nodes, and the clustering
coefficient is larger than that of a reference random network having
the same number of nodes and edges as the real network.

We also notice the small-world effect in the four FTNs. On one hand, the
average path length {\it L} and the diameter {\it D} are small. For the
FTNs, {\it L} ranges from 2.307 to 2.642, while the values are
between 2.774 and 2.933 in the corresponding equivalent random
networks generated by the Erdos-Renyi model with the same parameters
from each FTN. The diameters {\it D} in the FTNs are 5 or 6, while 3
or 4 in the corresponding random networks. On the other hand, these
networks are highly clustered. The clustering coefficients {\it C}
are 0.0480, 0.0453, 0.0499 and 0.0370 for FTN-all, FTN-met, FTN-rub
and FTN-oil, respectively,  and the value of each FTN is larger than
that of the corresponding random network (with values of 0.0016,
0.0011, 0.0022 and 0.0012, respectively). These results are presented
in~\autoref{tab:swe-ftn}.

\begin{table}[!hbp]
    \captionstyle{indent}
    \caption{Small-world effect is seen in all the four futures trading
    networks with relatively large clustering coefficient {\it C} and small average path
length {\it L}, contrary to the corresponding random networks
generated by the Erdos-Renyi model with the same parameters (total
number of nodes {\it N}, total number of edges {\it E} and average
number of edges per node $<k>$) of the futures trading networks.}
    \label{tab:swe-ftn}
    \centering
    \begin{tabular*}{\textwidth}{@{\extracolsep{\fill}}  c *{7}{c}}
        \hline
        Network &\itshape N &\itshape E & $<k>$ & $k_{max}$ &\itshape C &\itshape L &\itshape D \\
        \hline
        \itshape FTN-all & 100994 & 8068676 & 159.8 & 36878 & 0.0480 & 2.470 & 6 \\
        \itshape random & - & - & - & 218 & 0.0016 & 2.774 & 4 \\
        \itshape FTN-met & 75262 & 3226119 & 85.7 & 19466 & 0.0453 & 2.642 & 6 \\
        \itshape random & - & - & - & 133 & 0.0011 & 2.905 & 4 \\
        \itshape FTN-rub & 55828 & 3364557 & 120.5 & 35694 & 0.0499 & 2.307 & 6 \\
        \itshape random & - & - & - & 169 & 0.0022 & 2.766 & 3 \\
        \itshape FTN-oil & 52208 & 1657268 & 63.5 & 23286 & 0.0370 & 2.494 & 5 \\
        \itshape random & - & - & - & 102 & 0.0012 & 2.933 & 4 \\
        \hline
    \end{tabular*}
\end{table}

\begin{figure}[!htp]
    \centering
    \includegraphics[width=\textwidth]{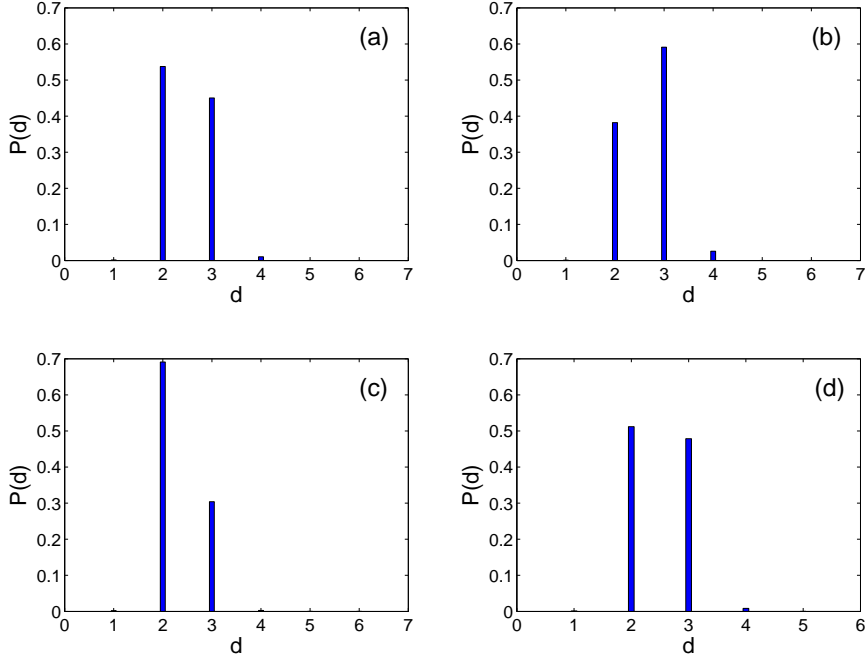}
    \caption{The shortest path distributions $P(b)$ of the four futures trading networks. Obviously, most of the shortest paths of each network are 2 or 3 in length. }
    \label{fig:path-distribution}
\end{figure}

The ratio of each FTN's clustering coefficient over that of its
corresponding random network is relatively small in comparison with
other real systems~\cite{Albert2002}. This implies no adequate
evidence of local cliquishness of a typical neighborhood within each
FTN, which may be due to the randomicity of matching counterpart
during trade execution. Furthermore, the small {\it L} and {\it D}
is because of 1) the existence of hub nodes (conforming to the large
maximum degree $k_{max}$ in ~\autoref{tab:swe-ftn}) that are bridges
between different nodes in the networks, and 2) new connections
generated between the existing nodes previously without direct
links, which provide more shortcuts to the networks.
The shortest path distribution
in~\autoref{fig:path-distribution} shows that most of the shortest
paths in the four networks are 2 or 3 in length, which helps
explaining the small-world effect.

\subsection{Hierarchical Organization}
To examine the hierarchical organization feature of a FTN, we check
{\it C(k)}, the average clustering coefficient of all nodes with the
same degree {\it k}. If {\it C(k)} follows a strict scaling law,
then hierarchical organization exists in the network. Some real
networks, such as the World Wide Web, the actor network and the Internet,
display hierarchical topologies~\cite{Ravasz2003}.

\begin{figure}[!htp]
    \centering
    \includegraphics[width=\textwidth]{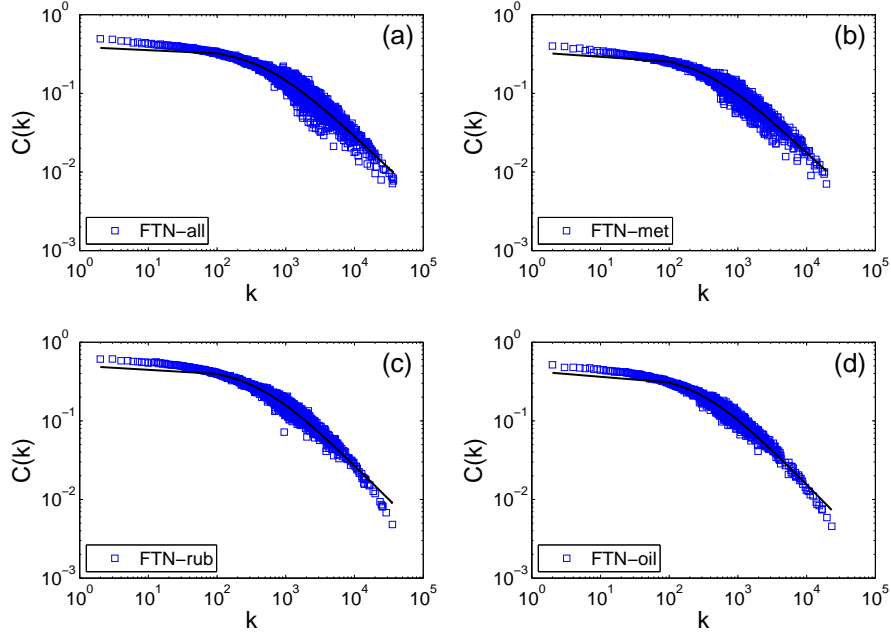}
    \caption{Clustering coefficient {\it C(k)} as a function $C(k) \sim (k+b)^{-\beta}$ of node degree {\it k} with nearly similar
    exponent of $0.87 \pm 0.03$ for the four futures trading networks, which implies that each network has a hierarchical architecture.}
    \label{fig:clustering-coefficient}
\end{figure}

For each FTN, as indicated in~\autoref{fig:clustering-coefficient},
the main part of {\it C(k)} obeys a scaling law of
\begin{equation} \label{e:CCD}
    C(k) \sim (k+b)^{-\beta}
\end{equation}
with nearly similar exponent $\beta=0.87 \pm 0.03$, which implies
that the network has a hierarchical architecture. Although the
majority of bargainers who make a few deals have a few links (low
{\it k}), most of these links join hub nodes that connect to each
other, resulting in a big {\it C(k)}. The high-{\it k} nodes are hub
bargainers, their neighbors that are low-{\it k} nodes are seldom
linked to each other, leading to a smaller {\it C(k)}. This implies
that ordinary investors are part of such clusters with high
cohesiveness and dense interlinking, and the hubs play a bridging
role to connect many separate small communities together into a
complete network. To some extent, the hub speculators' trading
behaviors flourish the futures market by improving market liquidity,
and consequently enhance the pricing function of the futures market.

\subsection{Betweenness Distribution}

Betweenness is a measure of the centrality of a node in a network;
it is also a measure of influence that a node has over the spread
of information in the network~\cite{Freeman1977}.

\begin{figure}[!htp]
    \centering
    \includegraphics[width=\textwidth]{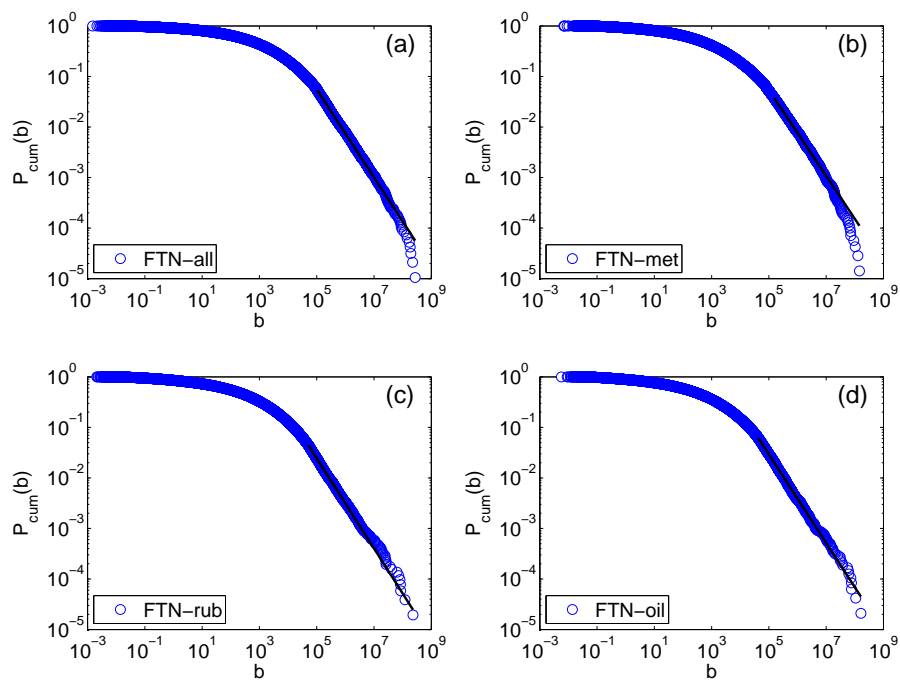}
    \caption{The cumulative betweenness distributions of the four futures trading networks, which follow a power-law function of $P_{cum}(b)\sim b^{-(\omega-1)}$.
    The exponents $\omega$ for FTN-all, FTN-met, FTN-rub and FTN-oil are 1.871, 1.850, 1.891 and 1.876, respectively.}
    \label{fig:betweenness-distribution}
\end{figure}

The betweenness of a given node is defined as the sum of the
fraction of shortest paths between all pairs of nodes in a network
that pass through the node. If there is more than one shortest path
between two nodes, each of which is counted into the total number
of shortest paths. To be precise, the definition is as follows:
\begin{equation} \label{e:CCD}
    b(i)={\sum_{j \neq k \neq i}{b_{jk} (i) \over b_{jk}}}
\end{equation}
where {\it $b_{jk}(i)$} is the number of geodesic paths from {\it j}
to {\it k} containing {\it i}, while {\it $b_{jk}$} is the total
number of geodesic paths linking {\it j} and {\it k}. In
~\autoref{fig:betweenness-distribution}, we present the cumulative
betweenness distribution function $P_{cum}(b)=\sum
_{b'=b}^\infty{P(b')}$ of nodes for the four FTNs. By exploring the
cumulative betweenness distributions $P_{cum}(b)$ of the four FTNs
with the method of Clauset et al.~\cite{Clauset2009}, like the degree
distribution, we find that they follow a power-law function, i.e.,
$P_{cum}(b)\sim b^{-(\omega-1)}$, corresponding to the betweenness
distribution $P(b)\sim b^{-\omega}$. The exponents $\omega$ for
FTN-all, FTN-met, FTN-rub and FTN-oil are 1.871, 1.850, 1.891 and
1.876, respectively. Furthermore, we also make a goodness-of-fit
test for each FTN and the results are shown in \autoref{tab:bd-fit-test}.
The {\em p}-value of the goodness-of-fit test for each power-law fit
of FTN indicates that the betweenness distribution of each network
is consistent with a power-law hypothesis.

\begin{table}[!hbp]
    \captionstyle{indent}
    \caption{ The estimations of the lower bound $b_{min}$ and the exponent $\omega$, and the goodness-of-fit tests of
    betweenness power-law behaviors of the four futures trading networks. The {\em p}-values indicate that
    the betweenness distribution of each network is consistent with a power-law hypothesis.}
    \label{tab:bd-fit-test}
    \centering
    \begin{tabular*}{\textwidth}{@{\extracolsep{\fill}}  c *{3}{c}}
        \hline
        Network &\itshape $b_{min}$ &\itshape $\omega$ &\itshape p-value \\
        \hline
        \itshape FTN-all & 103410.8 & 1.871 & 0.237 \\
        \itshape FTN-met & 149845.9 & 1.850 & 0.437 \\
        \itshape FTN-rub & 50322.7 & 1.891 & 0.334 \\
        \itshape FTN-oil & 42603.8 & 1.876 & 0.278 \\
        \hline
    \end{tabular*}
\end{table}

Information communication (e.g. money or asset transfer) exists in
FTNs, and money flow is considered as the index of financial markets.
The power-law distribution demonstrates that the nodes having high
betweenness own such a potential to control information flow passing
between node pairs of the network. Accordingly, the participants
with high betweenness centrality play decisive role in promoting
market boom and enhancing financial function.

\subsection{Disassortative mixing}

In a network, the preference for a node to attach to others that are
similar or different in some way is called assortativity, which is
often examined in terms of degree correlation between two nodes.
Such a correlation is captured by two prominent measures: the
assortativity coefficient~\cite{Newman2002,Newman2003b} and the
average nearest-neighbor degree~\cite{Pastor-Satorras2001}.

The assortativity coefficient is essentially the Pearson correlation
coefficient of degree between a pair of linked nodes, and is defined
as
\begin{equation} \label{e:AC1}
    r={1\over\delta_q^2} \sum_{jk}jk(e_{jk}-q_jq_k)
\end{equation}
where $q_k$ is the distribution of the remaining degree, i.e., the
probability that a node has {\it k} other edges, $q_j$ is the
probability that a node has {\it j} other edges, $e_{jk}$ is the
joint probability distribution of the remaining degrees of the two
nodes at either end of a randomly chosen edge, $\delta_q$ is the
standard deviation of the distribution $q_k$,
$\delta_q^2={\sum_kk^2q_k-{(\sum_kkq_k)}^2}$. The coefficient {\it
r} lies in the range $-1 \leq r \leq 1$. Positive $r$ values
indicate the existence of assortative mixing pattern, a preference for
high-degree nodes to attach to other high-degree nodes, while
negative $r$ values imply disassortative mixing, i.e., high-degree
nodes to connect, on average, to low-degree ones. Zero $r$ means no
assortativity.

For practical evaluation, the assortativity coefficient can also be
calculated by:
\begin{equation} \label{e:AC2}
    r={{M^{-1}{\sum_i j_i k_i}-[M^{-1}{\sum_i {1\over2} (j_i + k_i)}]^2} \over {M^{-1}{\sum_i {1\over2} (j_i^2 + k_i^2)}-[M^{-1}{\sum_i {1\over2}(j_i + k_i)}]^2}}
\end{equation}
where $j_i$ and $k_i$ are the degrees of the nodes at the ends of
the {\it i}th edge, with $i$=1...$M$, $M$ denotes the number of
edges in the network. We calculate the assortativity coefficients of
the four FTNs, and obtain the values of {\it r}, which are -0.1522,
-0.1590, -0.1594 and -0.1608 for FTN-all, FTN-met, FTN-rub and
FTN-oil, respectively. These negative values undoubtedly show that
these FTNs are disassortative mixing.

Another means of measuring the degree correlation is to examine the
property of neighbor connectivity, i.e., the average nearest-neighbor
degree $\langle k_{nn} \rangle$ of a node with degree {\it
k}, which is defined as
\begin{equation} \label{e:KNN}
    \langle k_{nn} \rangle=\sum_{k'}k'P(k'|k)
\end{equation}
where $P(k'|k)$ is the conditional probability that an edge points
from a node of degree {\it k} to a node of degree {\it k'}. If
$\langle k_{nn} \rangle$ increases with {\it k}, the network is
assortative since it implies that the nodes of high degree tend to
connect to the nodes of high degree. Conversely, if $\langle k_{nn}
\rangle$ decreases with {\it k}, the network is disassortative,
which means that nodes of high degree are more likely to have
nearest neighbors of lower degree. If $\langle k_{nn} \rangle$ is
independent of {\it k}, there is no correlation between nodes of
similar degree.

\begin{figure}[!htp]
    \centering
    \includegraphics[width=\textwidth]{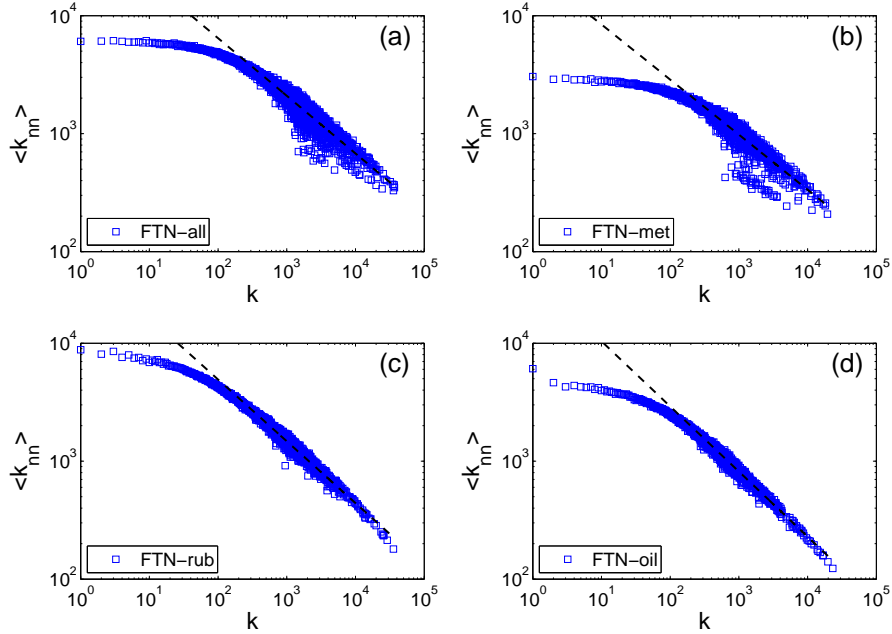}
    \caption{The average nearest-neighbor degree $\langle k_{nn} \rangle$
    as a function of the node degree {\it k} for the four futures trading networks, with the slope of value -0.52$\pm$0.03}
    \label{fig:degree-knn}
\end{figure}

In~\autoref{fig:degree-knn}, we plot the neighbor connectivity to
depict the overall assortativity trends of the four FTNs. We can see
that the FTNs show a power-law dependence on the degree $\langle
k_{nn} \rangle \sim k^{\mu} $, with similar exponent
$\mu=-0.52\pm0.03$. The result of $\langle k_{nn} \rangle$
decreasing with {\it k} clearly infers the existence of
disassortativity in the FTNs.

The disassortative mixing of FTNs discloses that their
assortativities are similar to that of technological and biological
networks, but are different from social networks. The nontrivial
correlation property of the FTNs indicates that the active
speculators with high degree do not frequently trade with each
other, on the contrary, they tend to make deals with seldom-trading
participants. This negative correlation suggests to monitor fluidity
risk of the futures market by tracking the active speculators.

\subsection{Power-Law Evolution}
Up until now we have explored topological properties of static FTNs.
Here we will consider some dynamic characteristics of the FTN-all
network by checking its evolutional process, which can characterize
the real trading behavior of the futures market.

\begin{figure}[!htp]
    \centering
    \includegraphics[width=\textwidth]{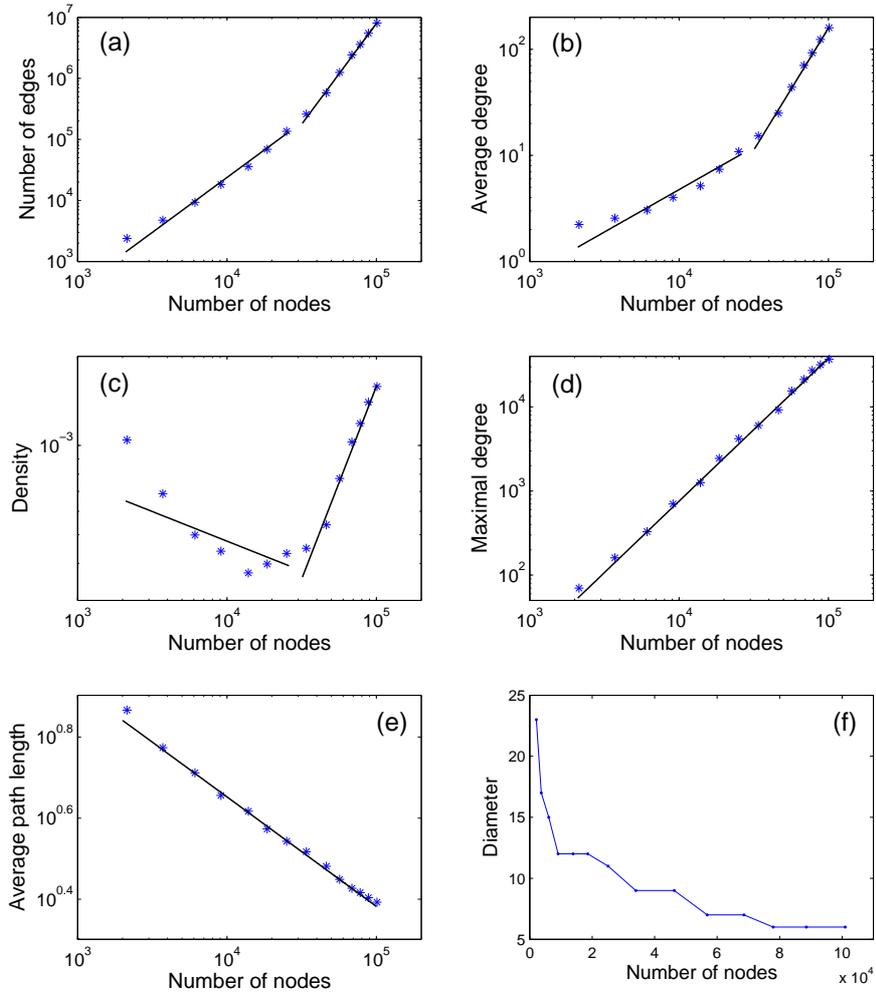}
    \caption{(a) The number of edges versus the number of nodes in log-log scale for
    FTN-all; a power law is observed for the slopes of 1.8 and 3.3, respectively in the evolution process.
    The critical point is at the end of the first trading day.
    (b) The average degree super-linearly grows in a power law with the slopes of 0.8 and 2.3, respectively.
    (c) The density changes over time in an approximate power law with slopes of -0.2 and 1.3,
    which account for the observation that the curve first falls and then rises.
    The conversion of edge growth's driving factors results in the different slopes in the power law underlying the network's evolution.
    (d) Maximum degree grows in a power law with regard to the network size, and the slope is 1.7.
    (e) The average path length decays in a power law with a slope of -0.27.
    (f) The diameter decreases as the number of nodes increases.}
    \label{fig:eresults}
\end{figure}

Like many real-life
systems~\cite{Faloutsos1999,Broder2000,Jeong2000}, FTN-all also
exhibits accelerated growth~\cite{Dorogovtsev2001} as shown
in~\autoref{fig:eresults}(a), which manifests that the number of
edges increases faster than the number of nodes. Furthermore, the
relationship between the number of edges and the number of nodes in
FTN-all follows a power-law function $e(t)\sim n(t)^ {\alpha}$.
Here, {\it e(t)} and {\it n(t)} respectively denote the numbers of
edges and nodes of the network at time {\it t}, and $ \alpha$ is the
exponent. Such a relationship is termed the densification power law by
Leskovec et al.~\cite{Leskovec2007}. We can see that the fitting
curve in ~\autoref{fig:eresults}(a) consists of two segments with
different slopes, 1.8 and 3.3, respectively. We notice
that the critical point of the broken fitting line is located in the
place where the network size is about 30000, and its corresponding
time is exactly the end of the first trading day, which means that
the mechanism of intraday growth of edges in FTN-all is totally
different from that of the following days.

Besides, we observe that the average degree and density of FTN-all
super-linearly grow with the network size, and both have broken
lines, as shown in~\autoref{fig:eresults}(b) and (c). The changes in
both average degree and density over time are directly related to
the change of edges' number. The average degree $\bar k(t)$ and
density {\it d(t)} are respectively evaluated by $\bar k(t)={2e(t) /
n(t)}$ and $d(t)={2e(t) / n(t)(n(t)-1)}$, and thus we obtain $\bar
k(t)\sim n(t)^ {\alpha-1}$ and $d(t)\sim n(t)^ {\alpha-2}$. In
~\autoref{fig:eresults}(b), the two slopes of the broken line are
0.8 and 2.3, respectively, and in~\autoref{fig:eresults}(c) they are
-0.2 and 1.3, respectively, which explains why the broken line of
the density first falls and then rises. According to the empirical
viewpoint of Leskovec et al.~\cite{Leskovec2007}, the changing trend
of the average degree of FTN-all in ~\autoref{fig:eresults}(b) suggests
that the network becomes denser as network size increases. However,
from the viewpoint of network density based on the benchmark of
complete graph, the curve in ~\autoref{fig:eresults}(c) implies that
FTN-all first becomes sparser and then quickly turns denser as
network size expands.

\autoref{fig:eresults}(d) shows that the maximum degree $k_{max}(t)$
of FTN-all grows in power law: $k_{max}(t)\sim n(t)^ {\beta}$ with
$\beta$=1.7. This observation indicates that the active participants
in the futures market keep vigorous all the time, and the day
traders (those who frequently buy and sell futures within the same
trading day so that all positions will customarily be closed before
the market closes on that day) constitute the core part of the
active participants.

In some real complex networks~\cite{Watts1998,Albert1999}, the
average path length scales logarithmically with the number of nodes,
and the diameter slowly grows with the network size. However, in
FTN-all we notice that the average path length decreases as the
number of nodes increases, so does the diameter. The average path
length decays in a power law with a slope of -0.27 in
~\autoref{fig:eresults}(e), while the diameter of FTN-all decreases
as shown in~\autoref{fig:eresults}(f). As the network size grows,
the simultaneous shrinkage of both the average path length and the
diameter in FTN-all provides another evidence of densification in
the network.

The exposed properties above indicate that the nodes in FTN-all
become closer to each other as the network size grows. These
properties indicate the real trade behavior of the futures
market. In the futures market, deals come from two sources: the
trading by new investors who have just entered the market, and the
transactions between the existing speculators. Each new investor
makes a few deals for the sake of caution, while the existing
speculators, especially the day traders, make a large number of
transactions. The latter contribute many deals between the
participants who have no trading relationship previously, which
results in new links between the nodes that have no direct
connections previously in FTN-all, and consequently accounts for the
properties mentioned above. Additionally, the different
dominant sources of edge addition during network expansion cause the
discontiguous slopes of two different fitting lines before and after
the switching point, happening at the time when the network size is about
30000 in each logarithmic plot of ~\autoref{fig:eresults}(a)-(c),
and exactly corresponding to the end of the first trading
day. In the first trading day, almost all investors are newcomers,
so the edge growth originating from new investors is the dominant
factor. After that day, the number of new investors and their
transactions tend to remain steady, while the edges generated among
the existing participants increase rapidly and outnumber those
generated by the new investors. The switch between the edge growth's driving
factors explains the existence of two different exponents in the
power law underlying the network's evolution.

\section{Conclusion and Future Work}
\label{sec:conclusion}

In this paper, futures trading networks have been extensively analyzed
from the perspective of complex networks. We constructed FTNs by
using real trading records covering three months' operation in the
Shanghai Futures Exchange. We found a number of interesting
statistical properties of the networks, including scale-free
behavior, small-world effect, hierarchical organization, power-law betweenness distribution,
disassortative mixing and power-law evolution. Some unique features
of the FTNs are possibly due to the randomicity of matching
counterpart during deal execution and the continuous generation of new
links between previously unconnected nodes, which play an important
role in determining the structures of futures trade networks.
Although three small networks (or sub-networks) are derived from a
partition of the whole market dataset, all their statistical
properties are consistent with that of the whole market network.

As only undirected graphs are considered in this paper, it is worth
further studying the directed FTNs based on trading direction,
such as from buyer to seller, which reflects the money flow in the
futures markets. In Section~\ref{subsec:scale-free}, we simply
established the weighted FTNs to demonstrate that active market
participants get more active, but it is not enough. Actually, we can
construct weighted FTNs or weighted and directed FTNs to investigate
the market behaviors of active participants. Moreover, the
generation model of FTNs is also an important issue for further
exploration, as it can provide valuable insight on financial trade
monitoring and risk control.

\section*{Acknowledgements}
We thank Prof.~Yunfa Hu and Dr.~Zhongzhi Zhang for their valuable
comments and suggestions. This work was supported by the National
Natural Science Foundation of China under Grant numbers 60873040 and
60873070. Shuigeng Zhou was also supported by the Open Research
Program of Shanghai Key Lab of Intelligent Information Processing.
Jihong Guan was also supported by the ``Shu Guang" Program of
Shanghai Municipal Education Commission and Shanghai Education
Development Foundation.

\section*{References}
\bibliographystyle{elsarticle-num}
\bibliography{References}

\end{document}